# Search Engine Drives the Evolution of Social Networks

Cai Fu
Huazhong University of
Science and Technology,
Wuhan, China
fucai@126.com

Chenchen Peng
Huazhong University of
Science and Technology,
Wuhan, China
pcclucky@163.com

Xiao-Yang Liu
Shanghai Jiao Tong University,
Shanghai, China
xl2427@columbia.edu

ABSTRACT
The search engine is tightly coupled with social networks and is primarily designed for users to acquire interested information. Specically, the search engine assists the information dissemination for social networks, i.e., enabling
users to access interested contents with keywords-searching and promoting the process of contents-transferring from the source users directly to potential interested users. Accompanying such processes, the social network evolves as new links emerge between users with common interests. However, there is no clear understanding of such a \chicken-and-egg" problem, namely, new links encourage more social interactions, and vice versa. In this paper, we aim to quantitatively characterize the social network evolution phenomenon driven by a search engine. First, we propose a search network model for social network evolution. Second, we adopt two performance metrics, namely, degree distribution and network diameter. Theoretically, we prove that the degree distribution follows an intensi_ed power-law, and the network diameter shrinks. Third, we quantitatively show that the search engine accelerates the rumor propagation in social networks. Finally, based on four real-world data sets (i.e., CDBLP, Facebook, Weibo Tweets, P2P), we verify our theoretical _ndings. Furthermore, we _nd that the search
engine dramatically increases the speed of rumor propagation.

## 1. INTRODUCTION

The search engine is primarily designed for users to acquire interested information in social networks. Recent years, the search engine is widely used by social network users. As the number of users increasing, the search engine is becoming more tightly coupled with social networks. A recently released statistic shows that as of the second quarter of 2016, 1.13 billion daily active users visited the Facebook web site [1], with a 45% growth compared with that of 2012. The increasing number of nodes indicates that the social network structure evolves over time [2, 3, 4].

As the search engine assists the information dissemination of social networks, we notice that the search engine plays an important role on the network evolutionary process. It enables users to access interested contents with keywords-searching and promotes the process of contents-transferring from the source users directly to potential interested users. Accompanying such processes, the social network evolves as new links emerge between users with common interests. We are interested in the following questions:

*How does the search engine drive the social network evolutionary process? What are the quantitative effects caused by the search engine?*

To address these issues, we are faced with the following three major challenges:

_ The search engine accelerates the information dissemination of social networks thus driving the evolution of social networks. Both the page ranking process utilized by the search engine and users' information accessing behaviors complicate this evolutionary process.

_ As the social network topology changes dynamically, the "chicken-and-egg" problem arises: new links encourage more social interactions, and vice versa.

_ To quantitatively characterize the evolutionary process, we need to properly model the network evolutionary process driven by the search engine and adopt basic metrics for network structure changes.

To characterize the evolutionary process driven by the search engine, we propose a `search network` model. In this model, the search engine effects the network evolutionary process by randomly adding new links between users. To quantitatively characterize the network evolutionary process, we adopt two metrics: degree distribution and network diameter. The network evolutionary process mainly involves the dynamic changes of edges and nodes, which is closely related with the degree distribution. Network diameter can be used to analyze the maximum information propagation delay between users who are interested in the same topics. Our model is based on the groundbreaking work of Silvio Lattanzi [7] that proposed bipartite models of social networks. At each time slot, a new user comes with a certain of probability and chooses a prototype by preferentially. Then the new user randomly copies edges from the prototype. Our model further considers that the search engine adds connections between users and topics, thus creating links between users and also between topics.

In this paper, the main contributions are as follows:

_ We reveal that the social network evolution is driven by the search engine. The search engine creates short-cuts between network nodes in a community, or two overlapping communities, or even two separated social networks. We propose a search network model that takes into consideration of both node-joining and link- adding.

_ We characterize how the search engine influences the metrics of network evolutionary process, such as degree distribution and network diameter. We mathematically and theoretically prove that the search network has an intensified power law degree distribution and network diameter significantly decreases.

Based on four real-world social network data sets, we verify our theoretical results. As an example, we show that as the search engine drives the social network evolution, the rumor/information propagation process is accelerated dramatically.

The rest of this paper is organized as follows. In Section 2, we present a search network model. The theoretical mathematical analyses of degree distribution and network diameter are presented in Section 3, respectively. Evaluations are given in Section 4 and related work is presented in Section 5. Finally, we conclude our work in Section 6.

## 2. NETWORK MODEL

We first introduce a search network model with a detailed algorithm. Then, we present two important metrics of network evolutionary process, i.e., degree distribution and network diameter.

### 2.1 Search Network Model

Our search network model consists of three entities: users, topics, and a search engine. We describe the evolutionary process as follows:

*1) User-topic graph:* As is shown in Fig. 1(a). We use $B(U, T)$ to represent the search networks, where U denotes the set of users and T represents the set of topics. In this graph, an edge exists between a user and a topic if the user is interested in the topic. Some of the edges are added by the search engine. Generally, users are connected with topics according two mechanisms: "preferentially chosen" and "edge copying".

*2) Preferentially chosen:* When a user comes, it preferentially chooses a prototype among exisiting users with probability proportional to their degrees. As shown in Fig. 1(b), the last user has the highest degree and is most likely to be chosen as the prototype. Here we choose the last user to be the prototype and use the gray lines to represent the edges that will be copied.

*3) Edge copying:* Then follow the "edge copying" step. The new user picks an exisiting user as its prototype and copies its edges (according to some probabilistic model). As shown in Fig. 1(c), since the prototype is connected with $topic1; topic2; topic3$ at the same time, the new user will establish connections with $topic1; topic2; topic3$ with a certain of probability. Here the new user is connected with $topic2, topic3$ (the red lines).

Network evolves over time and we focus on how the search engine influences the network evolutionary process. As shown in Fig. 1(d), besides $topic2; topic3$, the user may also be interested in $topic1$. Then it uses search engine to search for $topic1$. The searching process establishes indirected edge between the user and $topic1$ (the green dotted line) and a new affiliation between them is set up. Since the evolutionary process of a new topic is a symmetrical process of a user, we omit it here.

We present the evolutionary algorithm in Algorithm 1 to describe the search network evolutionary process. In the algorithm, the most important step is that the new user will be connected with several topics by the search engine with probability $p_t$. Moreover, $p_t$ is related with the ranking algorithm of the search engine and the similarity between users and topics. When there is no search engine, this step will be skipped.

In order to understand the intuition behind the evolutionary process of search networks, let us consider, for example, the citation graph among papers. When an author writes a new paper, he probably has in mind some older papers. One of the older papers will be the prototype, and he is likely to write on (a subset of the) topics considered in this prototype. Moreover, the author also writes some novel knowledge and searches for other topics, then the paper is connected with other topics by the search engine. Similarly, when a new topic emerges in the literature, it is usually inspired by an existing topic (prototype) and it probably has been foreseen by older papers. Novel characteristics of the topic will also attract other papers which are interested in the novel topic. They use the search engine to search for the topic.

---

**Algorithm 1** Evolving Search Network $B(U, T)$

---

Fix two integers $e_u, e_t > 0$, and let $\beta \in (0, 1)$, $p_t \in (0, 1)$.
Process:
**At time 0:**
  Bipartite graph $B(U, T)$ is a simple graph with at least $e_u \, e_t$ edge, where each node in U has at least $e_u$ edges and each node in T has at least $e_t$ edges; $p_t$ represents the probability of that a user links to topics in $B(U, T)$ at time t by the search engine.
**At time t>0:**
  **(Evolution of U)** With probability $\beta$:
   **(Arrival)** A new user q is added to U.
   **(Preferentially chosen prototype)** A node $q' \in U$ is chosen as *prototype* with probability proportional to its degree
   **(Edge copying)** $e_u$ edges are "copied" from $q'$; that is, $e_u$ neighbors of $q'$, denoted by $u_1, ..., u_{e_u}$ are chosen uniformly and the edges $(q, u_1), ..., (q, u_{e_u})$ are added to the graph.
   **(Search engine adding edges)** Then, q connects with several topics by the search engine with $p_t$.
  **(Evolution of T)** With probability $1 - \beta$, a new topic u is added to T following a symmetrical process, adding $c_t$ edges to u. Then it connects with several users by the search engine.

---

### 2.2 Network Metrics

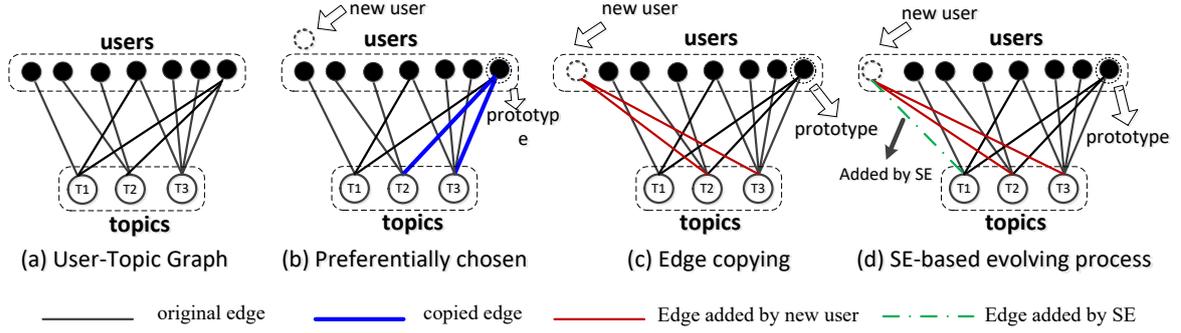

Figure 1: Illustration of the user-topic graph and the evolving process. (a) User-User Graph. (b)~(d) Evolving steps of the search network.

We adopt two metrics to characterize the network structure changes, i.e., degree distribution and network diameter. First, the network evolutionary process mainly involves the dynamic changes of edges and nodes. These changes are closely related with the degree distribution. Second, network diameter can be used to analyze the information propagation between users who are interested in the same topics. So it is also closely related with the network evolutionary process.

*1) Degree Distribution:* A network consists of nodes and edges. For each node, its degree is the number of edges that coming out from the node. Degree Distribution is a probability distribution of the nodes' degrees. The previous work in [7] noticed that the degree distribution of Internet graph obeys "power law", that is, for some constant $\alpha > 0$, the fraction of nodes of degree $d$ is proportional to $d^{-\alpha}$. Similarly, our model also includes the "preferentially chosen" and "edge copying" steps. These two steps guarantee a "power law" degree distribution. The only different part is that the search engine makes more connections between nodes. Obviously, the degrees of nodes have an increasing trend compared with the no-engine state. Consequently, the "power law" is intensified in search networks.

*2) Network Diameter:* The diameter of a network $Ds$ is the maximum of all shortest distance $d_{ij}$, namely $Ds = \max_{1 \leq i,j \leq N} d_{ij}$, where $i, j$ represent two different nodes in the network. Network diameter is used to analyze the information propagation which is the key function of social networks. Nodes gather together by the same interest. This process makes the information propagation regionalization. In search networks, the nodes can search information across different regions. As a result, the propagation speed is increased and the network diameter is decreased faster.

## 3. NETWORK PROPERTIES

We characterize the degree distribution and the network diameter. Theorem 1 states that the degree distribution is the summation of a series of power law distributions. We analyze the probability that a user links to topics by the search engine. Then, we prove that the network diameter decreases significantly.

## 3.1 Degree Distribution

We start by showing the degree distribution in evolving search network model $B(U, T)$. Once a new node is added to $T$, it adds at least $c_t$ edges with end points in $U$, which results in the change of the degree distribution in $U$. We have the symmetric situation for new nodes in $U$.

For convenience, Table 1 denotes all notations that will be used in later analyses, proofs and discussions. We first introduce several useful lemmas:

*Lemma 1.* If a sequence $a_t$ satisfies the recursive formula $a_{t+1} = (1 - b_t/t)a_t + c_t$ for $t \geq t_0$, where $\lim_{t \to \infty} b_t = b > 0$ and $\lim_{t \to \infty} c_t \geq c$ exists. Then $\lim_{t \to \infty} a_t/t$ exists and equals $c/(1 + b)$.

*Lemma 2* [7]. For the bipartite graph $B(U, T)$ generated after n steps (no engine), the degree sequence of nodes in U follows a power law distribution with exponent $\alpha = -2 - \frac{c_u \beta}{c_t(1-\beta)}$, when $n \to \infty$.

*Theorem 1.* For the bipartite graph of evolving search network $B(U, T)$ generated after n steps, when $n \to \infty$, the degree of nodes in U is a sum of a series of power law distributions with exponent $\alpha = -2 - \frac{c_u \beta}{c_t(1-\beta)}$.

*Proof.*

For $B(U, T)$, the degree of a node in U can increase if the following two events happen:

_ a new node is added to T and connects with nodes in U automatically.

_ a new node is added to T and connects with nodes in U by using the search engine.

In evolving social networks, the degree of nodes changes over time, we let $EN_t^i$ denote the expected number of nodes in U of degree i at time t. We look into the cases for $i = c_u$ and $i > c_u$ respectively, in the following.

*Case of $i = c_u$.*

$$\lim_{t \to \infty} EN_t^i / t = \frac{(\beta - p_t u c_t)(c_u \beta + c_t(1-\beta))}{c_u \beta + c_t(1-\beta) + (1-\beta)c_u c_t}$$

*Case of $i > c_u$.*

$$\lim_{t\to\infty} EN_t^i /t \sim i^{-2-\frac{c_u\beta}{c_t(1-\beta)}} + p_t uc_t(\frac{c_u\beta}{c_t(1-\beta)} + 1) \sum_{k=c_t}^{i} k^{-3-\frac{c_u\beta}{c_t(1-\beta)}}$$

As has been proved, for the evolving search networks, the degree distribution is a summation of a series of power law. Moreover, the search engine not only increases the degree of nodes with high degree value, but also greatly increases the value of $c_t$. Because the nodes with high degree value just like hot topics, they can attract relatively more connections under the "preferentially chose" mechanism. At the same time, the lowest degree $c_t$ will also be increased by the search engine edge-increased process. The above proofs prove that the power law degree distribution is intensified in the search networks.

## 3.2 Network Diameter

For a search network, studying its shortest path problem has considerable signification. We can analyze information propagation by the shortest path. In traditional networks, two nodes in two separated groups can not share information. In search networks, the search engine creates a virtual edge, which links these two separated groups. Then the information can spread in these two separated groups and most nodes in these two groups can share information. Therefore, search engine cuts down the length of path.

Another important network characteristic is network diameter, which is closely related with the shortest path. We calculate the trend of the network diameter to observe the effect of search engine. The expected value of the network diameter in search networks is lower than that without search engine. In an extreme condition, that is, a user connects all topics by the search engine, the network diameter will be close to 1. And any two users can share information.

In a bipartite graph of a social network, the worst network diameter is $D_{s1} = u + 1$, where $u$ is the number of topics in $B(U, T)$. In a bipartite graph of a search network, the worst network diameter is $D_{s2} = u - p_t u + 1$. Just compare the upper limit of network diameter, we can obtain $D_{s1} > D_{s2}$. So we come to a preliminary conclusion that the search engine shrinks the network diameter and the rigorous certification follows.

The expected value of the shortest path of a social network is defined as follows:
$$E_{route} = \sum (\text{the number of certain kind of route}) \cdot (\text{the probability of this route}).$$

We define that an edge between two nodes is a unit route. When a new user is added to U, new connections between the new user and topics will be established with a certain of probability. Let $E_0$ denote the number of nodes which have edges at current, $n$ is the total number of nodes in U. If an edge is exactly exist, we consider the probability of this edge is 1, or if the edge is added with some probability, we use a certain variable (from 0 to 1) to represent the probability of edge. In order to simplify the certification, we consider that, the probability of edges between two separated nodes is the maximum of the shortest distance (denoted by $D_{MAX}$, $D_{MAX} \gg 1$). We ignore the edge overlapping situation to get a overview of the network shortest path.

For a search network, at time slot t, the search engine makes a new user connect with topics with a certain probability $p_t$. We have that
$$E_{route}1 = E_0 \cdot 1 + c_u\beta + c_t(1-\beta) + p_t uc_t + D_{max}(\frac{n(n-1)}{2} - E_0 - c_u\beta - c_t(1-\beta) - p_t uc_t)$$
$$= E_0 + D_{max}(\frac{n(n-1)}{2} - E_0) - (D_{max} - 1)(c_u\beta + c_t(1-\beta) + p_t uc_t). \quad (3)$$

For a social network without the search engine, the influence of the search engine can be ignored, thus the expected value of the shortest path as follows:
$$E_{route}2 = E_0 \cdot 1 + c_u\beta + c_t(1-\beta) + D_{max}(\frac{n(n-1)}{2} - E_0 - c_u\beta - c_t(1-\beta))$$
$$= E_0 + D_{max}(\frac{n(n-1)}{2} - E_0) - (D_{max} - 1)(c_u\beta + c_t(1-\beta)). \quad (4)$$

From (3) and (4), we can easily know that $E_{route}2 > E_{route}1$. The result shows that the search engine shrinks the network shortest path. In search networks, besides the tradition evolutionary process, nodes can also be connected by the search engine with a certain of probability. Thus the probability that a route is created between two nodes will be increased. It results in the decrease of the separated nodes' number. So the value of the maximum shortest path will be decreased. Therefore, in search networks, the expected value of the network diameter will be shorten.

## 4. EVALUATION

In this section, we first conduct an experiment to verify our derivation that in search networks the degree distribution is a sum of a series of power law distributions. Then, we conduct the second experiment to show that the network diameter decreases significantly. Finally, we design a rumor propagation experiment and prove that the spread of rumor is much faster in search networks.

## 4.1 Experiment Data Sets

The degree distribution experiment and rumor propagation experiment are carried out based on four real data sets. These data sets are used as initial state. Every time slot we add a new node to the data set according to the evolutionary algorithm proposed in section 2. Then we keep tracing the change of network properties during the evolutionary process. By this method we can make our results well confirm to the real-world social networks. CDBLP data set represents a social network of the academic field. Facebook is a prevalent social network that is used to chat with others. Weibo is a popular social network that is used to share information with others. Most people use the mobile Weibo to follow the information by the smart phone. Weibo data set is also a mobile social network. P2P is a traditional social network that is used to share files. The emulational dataset is a general social network for comparing the propagation effect without the search engine.

*CDBLP data sets.* CDBLP was published by "Automation Discipline Innovation Method" research topic of Chinese Academy of Sciences Institute of automation and it derived from the network of Computer Chinese Journal. By analyzing this data set, we can supply the propagation effect of the search engine for a social network of professional field. The part of the data on 2010 is used in our experiment. The total number of CDBLP nodes is 4000.

*Facebook data sets*. Facebook was published by the SNAP library. Since the Facebook data set is one of the biggest social network web sites. By the experiment of Facebook data set, we can analyze the propagation effect of typical social network. Many users use the Facebook to chat with other people and share interest topics. We use the proximate Facebook data set to test. The data on the 2012 is used in our experiment. The total number of the Facebook nodes is 10000.

*Weibo data sets*. Weibo was published by the zhyoulun's blog. Since Weibo is one of the biggest mobile social sites. By the experiment of theWeibo data set, we can analyze the influence of the search engine in the mobile social network. In recent years, most users use the mobile weibo to follow others' news. The data on the 2015 is uses in our experiment. The total number of the Weibo nodes is 100000.

*P2P data sets*. A sequence of snapshots of the Gnutella peer-to-peer _le is the earliest social network for sharing network from August 2002. Since the P2P is a tradition social network that is used to transmit and share files. We can know the influence of the search engine for the fixed social network. There are total of nine snapshots of Gnutella network collected in August 2002. Nodes represent hosts in the Gnutella network topology and edges represent connections between the Gnutella hosts. Some nodes gather together with a certain clustering coefficient to share a type of file.

## 4.2 Degree Distribution Experiment

We have proved that the degree distribution is a summation of a series of power law distributions in search networks. Compared with the degree distribution of evolutionary networks without the search engine, the power law degree distribution in the search networks is more significant. We design an experiment to verify the above views.

Starting from initial time slot $t = 0$, we set a unit time as the time interval and observe the degree distribution of evolutionary networks. Every time slot, a new user is added with a certain of probability. Then the new user is connected with topics by preferential attachment and edge coping. In the search networks, more connections may be built, that is, the new user would be connected with more topics by the search engine with probability $p_t$. The probability $p_t$ is affected by the similarity value and rank algorithm of the search engine. It is emphasized that the value of $p_t$ will not influence the overall trend of the degree distribution, and here we take 0.1 as its value. The experiment results show that different data sets do not change the overall trend of degree distribution and do not influence our final conclusion. The experiment results of different data sets are shown in Fig. 2(a)~(d) respectively. Since in real world networks only a few nodes may have degree below 10, so we start our records from degree 11.

Fig. 2 shows the degree distributions of the networks with or without the search engine. We can see that the trends of the degree distributions are the same. That is, the number of low-degree nodes is greater than that of high-degree nodes. To make our analysis more intuitive, we redraw the graph in a logarithmic coordinate as shown in Fig. 3. Since the power law distribution has a characteristic that when k and P(k) (or the number of nodes with degree k) both take the logarithm, the graph will show as a linear line. So we take advantage of this characteristic. The results prove that the degree distributions of social networks with or without the search engine both comply with power law. By comparing the two figures, we find that the degree distribution in search networks is more precipitous than that without the search engine. This phenomenon verifies our opinion that the search engine intensifies the power law. Also it seems that a series of power law distributions are superimposed to form the curve of degree distribution in search networks.

## 4.3 Network Diameter Experiment

We have proved that the network diameter decreases significantly in the search networks. We design an experiment to compare the network diameter of evolutionary networks with/without the search engine. In order to measure the network diameter more easily, we ignore the unconnected graphs and build a connected bipartite graph of social networks. In the connected graph, all nodes can connect to each other directly or by middle nodes.

We start from a connected graph of 10 users and 10 topics. Every time slot a new user is added with a certain of probability. Then the new user is connected with topics follows the evolutionary algorithm proposed above. When there is no search engine, topics are selected only by "preferential attachment" and "edge coping". In the search networks, the new user would be connected with more topics by the search engine with probability $p_t$. $p_t$ will not influence the overall trend of the network diameter and we take 0.1 as its value. We track the change of network diameter and record it at each time slot. The experiment results are shown in Fig. 4.

As shown in Fig. 4, except the beginning period of time, the search networks' diameter is always lower than the other. By analyzing the results, we come to a conclusion that the search engine cuts down the network diameter. As an intuition of the decrease of the network diameter, in the search networks the users contact more tightly with each other. The propagation of information between users becomes faster and wider.

## 4.4 Rumor Propagation Experiment

Search engine makes more connections and communications between users. The propagation of information is faster and wider. Rumor, a common kind of information, also takes advantage of search engine to accelerate its propagation in social networks. In this section, we study the rumor propagation in search networks according to the SIR model [4]. A user is either susceptible meaning that he/she has not yet received a particular item of rumor, or infectious meaning that he/she is aware of the rumor and is capable of spreading it to his/her contacts, or recovered meaning that he/she is no longer spreading the rumor. We quantify the rumor propagation effects in search networks in terms of rumor coverage.

### 4.4.1 Rumor Coverage

In order to analyze the rumor propagation effects in search networks, we take rumor coverage as the parameter. By comparing the trend of rumor coverage with/without the search engine, we can prove that the search engine accelerates the rumor propagation process.

Specifically, we calculate the rumor coverage $I_t$ as follows:

$$I_t = \frac{m_t}{n},$$

where $m_t$ is the number of nodes that are aware of the rumor on time slot $t$, and $n$ is the total number of nodes.

### 4.4.2 Parameter Settings

*Without the search engine.* Starting from initial time slot $t = 0$, we set a unit time as a time interval and observe the rumor propagation process. The rumor coverage of a time interval measures average propagation velocity of a unit time. Initial number of nodes that are aware of the rumor is $I_0 = 0.01n$ in each data set's experiment. The $n$ is the total number of nodes in each data set. A node can be aware of a rumor by the adjacent infected nodes with a certain probability $\lambda=0.1$ on the next time interval. Moreover, nodes can stop spreading the rumor with a certain probability $\mu=0.01$. We observe the rumor coverage during 200 time slots.

*With the search engine.* All parameters are set the same as the case without the search engine except the following part: a node can be aware of a rumor through the search engine with a certain probability $\xi=0.1$. We keep observing the propagation of rumor until the rumor coverage come to a stable state.

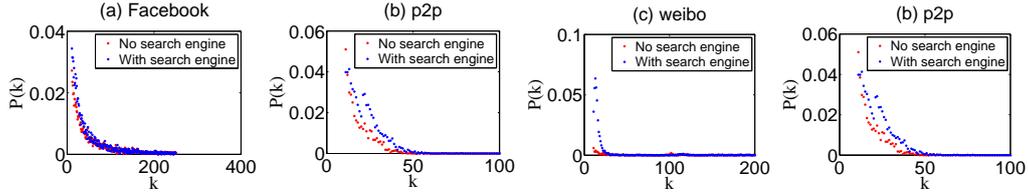

Figure 2: Degree distribution with/without search engine in different data sets.

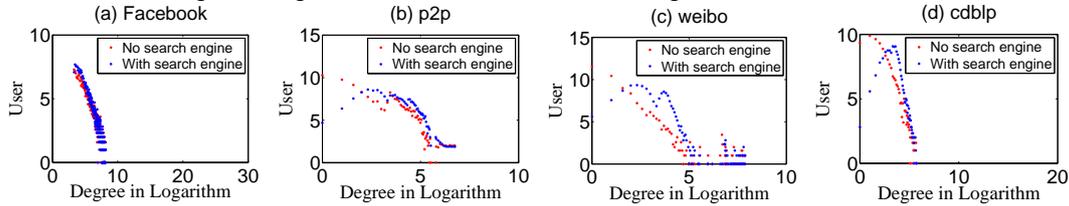

Figure 3: Degree distribution with/without search engine in logarithmic coordinate in different data sets.

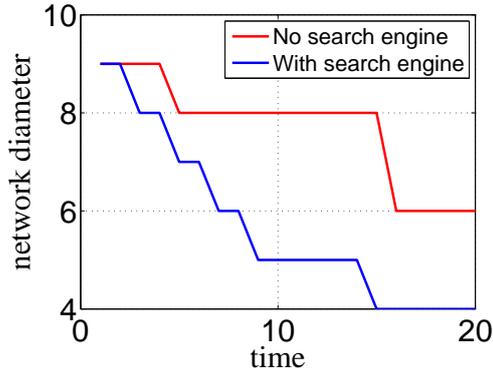

Figure 4: Network diameter.

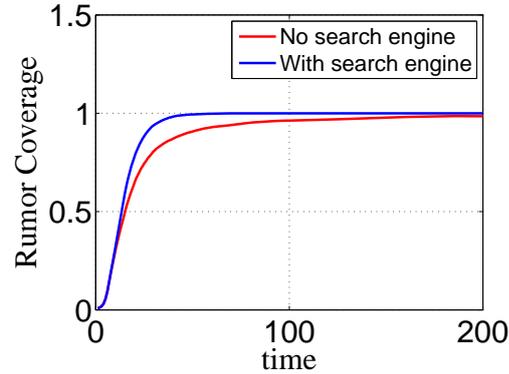

Figure 5: Rumor coverage.

### 4.4.3 Result Analysis

We use four real world data sets to compare the rumor coverage with/without the search engine. The experiment results show that different data sets do not change the over- all trend of rumor coverage and do not influence our final conclusion, so we do not list all the graphs here. The experiment results of Facebook data set are shown in Fig. 5.

As shown in Fig. 5, the trends of rumor coverage with or without the search engine both appear to rise significantly as the time goes and arrive a stable level in succession after a period of time. At the beginning, the two rumor coverage curves almost overlap. Then at time 10, two curves begin to separate. In the next time, the rumor coverage in the search networks is always larger and the curve becomes more precipitous than that without the search engine. When the rumor coverage reaches a stable state, the rumor coverage in the search networks is higher than the other, and they both appear to close to 1 (which means nearly all of the nodes get to know the rumor).

Thinking about the graph, we find that the search engine has a great impact on rumor propagation in social networks. Nodes get to know the rumor by the adjacent infected nodes or by the search engine. And finally the rumor coverage reaches a stable state. It represents that the rumor propagation reaches the maximize range. At the beginning, the rumor coverage curves almost coincide together, because the spreading area is not wide enough and the number of infected nodes is small, rumor propagation is hardly ever affected by the search engine. However, after time 10 the search engine begins to play an important role on the rumor propagation, the differences between two curves become obviously. Without the search engine, it is difficult for the rumor to spread across two separated groups. So the rumor propagation speed will be slower without the search engine. Finally, when both curves reach stable state, the rumor coverage in the search networks is higher. Because search engine makes more connections which further increases the rumor sources and propagation paths, so the rumor propagation is promoted.

In all, the search engine increases rumor sources and propagation paths. The search engine gathers rumors that are from the promulgator as rumor sources. The infected nodes will also create some outbound links in their own forum or other websites with a certain of probability. The search engine also gathers these links as novel rumor sources. Once the user clicks these rumor sources, the virtual propagation paths are established.

## 5. RELATED WORK

Recent years social network has become an important form of networks. Many kinds of network evolutonary process have been studied [14, 15, 16]. More and more social network users get used to acquire interested information by search engine. Antal van den Bosch et al. [17] conducted a 9-year longitudinal study to observe the search engine index size variability. Search engines bring too much convenience to a vast number of users and many technologies go hand in hand with the support of search engines [18, 19, 20].

With the development of search engine tools, the connections between network users are more closed and network structures become more complex. Search engine has always been updating to satisfy the increasing user needs. This phenomenn in turn leads to the network structures' change [21] [22]. As a result, the tradition network structures can not satisfy the new requirements any more. Many studies [2, 3, 4, 5] have paid out great efforts to describe the always-changing network structures. [23, 24] focus on the arrival and departure of users during the network evolutionary process. [25] studies node behaviors and predicts the link mechanism in coupled networks. Many models are proposed to analyze the complex network characteristics. Chung et al. [26] raise duplication models for biological networks. Ghoshal et al. [27] formulate a model which emphasizes the importance of the individual elementary mechanism. [28] introduces the public-private model of social networks. [29] studies network integration and proposes a energy-based model. However, none of them take search engine into consideration.

A suitable model is desired to describe the evolutionary process of search networks. In this paper, we propose a search network model. The model is rooted in sociology and leads to clean mathematical analysis as well as algorithmic benefits. The model develops from affiliation network. Affiliation network is certainly not new, it is based on a previous work of Sivakumar et al. [7]. In our model, users and topics are related by affiliation of the former in the latter. The evolutionary process of them follows "preferentially chosen" and "edge copying". Besides this two steps, users and topics are also related by the search engine with a certain of probability.

Nowadays, the research community overlooks the influence that the search engine has on social networks. If we could find the exact impact of search engine on network evolutionary process and analyze the relationship between search engine and network structure precisely and quantitively, we can take full advantages of search engine. Namely, we can maximize the benefits. What's more, studying the relationship between search engine and network structure can also help to find out the rules of information propagation (such as rumor propagation). It plays an important role on curbing the spread of the rumor. Combing our works with other current research results, we could even optimize the network structures and make the system develop in expected direction. The above is our final goal.

## 6. CONCLUSION

With the development of big data, more and more people use search engine to find information Great changes happen on the tradition network structures. In this paper, we build a simple bipartite graph model (the search network model) to describe the novel network structure. Based on the model, we simulate the evolutionary process of search networks. Then we prove that the degree distribution is a summation of a series of power law distributions and network diameter decreases significantly in search networks. Also experiments are conducted to verify the theory. Moreover, we design a rumor propagation experiment. The experiment results show that the search engine has great influences on the characteristics of social networks and accelerates the rumor propagation significantly.

In the future, we will continue to explore the exact impact of the search engine by studying more characteristics of social networks. We will also try to analyze the relationship between the search engine and network structure more precisely and quantitively. At the same time, we will take a further study on how to apply our research results to practical area, such as optimizing the network structure and making the system develop in our expected direction.